\documentclass[a4paper]{revtex4}
\usepackage{graphicx}

\begin{document}
\title{On Randomness in Quantum Mechanics}
\author{Alberto C. de la Torre }
\email{delatorre@mdp.edu.ar}
 \affiliation{Departamento de F\'{\i}sica,
 Universidad Nacional de Mar del Plata\\
 Funes 3350, 7600 Mar del Plata, Argentina\\CONICET}
\begin{abstract}
The quantum mechanical probability densities are compared with
the probability densities treated by the theory of random
variables. The relevance of their difference for the
interpretation of quantum mechanics is commented.
\end{abstract}
\maketitle
\section{INTRODUCTION}
Probabilities were introduced in quantum mechanics by Max
Born\cite{Born} when he proposed an interpretation for the wave
function $\Psi(x)$ of a particle as a probability amplitude.
Thereby $|\Psi(x)|^{2}$ is a probability density, that is, the
probability assigned to the event corresponding to the location
of the particle in the interval $[x,x+dx]$. Today, 80 years
later, we don't know the nature of this probability, that is, we
don't know whether this probability has an \emph{ontological} or
a \emph{gnoseological} character: we don't know whether the
particle \emph{is located} somewhere but we can not know where it
is, and the best that we can do is to give a probability for it,
or on the contrary, the very location of the particle \emph{is
diffuse} by nature. Although the Bell-Kochen-Specker
theorem\cite{bell2,koch} favours an ontological interpretation,
as can be easily seen in a recent illustration\cite{dlT} of the
theorem concerning position and momentum observables, the
opposite view, assuming context dependent definite values for the
observables, is not excluded. It soon became clear however that
these probabilities do not behave as such for mutually exclusive
events and the rule ``thou shall not add the probabilities but
their amplitudes instead'' was adopted although never understood.
A beautiful illustration of the necessity of this rule is given
by R. Feynman\cite{fey} in the analysis of the two slits
experiment.

Quantum mechanical ``probabilities'' are not added; however,
probabilities are studied by a rigourous mathematical theory of
random variables that tell us that, indeed, they \emph{can} be
added. Instead of requiring a change in the mathematical theory
we should perhaps accept that the quantum mechanical densities
\emph{are not probabilities}. In this work we will analyse the
quantum mechanical densities and compare them with the
probability densities defined in the theory of random variables.
We will see that, in general, the quantum mechanical densities do
not behave as the probability densities of random variables and
we will comment on the consequences of this for the
interpretation of quantum mechanics.

\section{RANDOM VARIABLES}
For completeness, we will recall some concepts of the theory of
random variables. A random variable $A$ is a mathematical entity
that may be assigned numerical values $a$ with some probability.
We can use the symbol $A\Rightarrow a$ to denote the event that
the random variable is assigned a value in the interval
$[a,a+da]$. We define the \emph{Probability Density Function}
$\varrho(a)$ as the probability of realization or appearance of
the event $A\Rightarrow a$. Of course, theses densities are
nonnegative and normalized.

Probabilities play an important role in classical physics and are
omnipresent in quantum mechanics so, in some cases, it is
interesting to consider the observables of a system as random
variables with their corresponding probability density functions.
For instance, we may study the possibility of assuming that the
position of a particle $X$ and its momentum $P$ are random
variables with their associated probability densities
$\varrho(x)$ and $\pi(p)$ that give the probability of location
of the particle and of having some value of momentum. Since
position and momentum are two independent concepts, in the sense
that a particle at a given position can have any value of
momentum, and some value of momentum can be realized at any
position, we can assume that position and momentum are described
by \emph{independent} random variables. We will later mention the
possibility that they are not independent.

Given a random variable $A$ with density  $\varrho(a)$, a
function $H(A)$ of the random variable $A$,  will have a density
$\eta(h)$ given by
\begin{equation}\label{functrv}
   \eta(h) = \int\!\!\!da\ \varrho(a)\ \delta\left(h-H(a)\right)\
   .
\end{equation}
This result is intuitive: the probability for some value $h$ is
the probability of the event $A\Rightarrow a$ times the
probability that $h=H(a)$, which is a sharp Dirac distribution,
added for al possible values of $a$. Similarly, it is easy to
understand that if $A$ and $B$ are independent random variables
and $\varrho(a)$ and $\pi(b)$ are the corresponding probability
densities associated with the events $A\Rightarrow a$ and
$B\Rightarrow b$, then the probability density $\sigma(c)$
assigned to the random variable $C=F(A,B)$ is given by
\begin{equation}\label{functrv1}
   \sigma(c) = \int\!\!\!da\!\!\!\int\!\!\!db\ \varrho(a)\ \pi(b)
      \ \delta\left(c-F(a,b)\right)\ .
\end{equation}
This result can not be avoided. If by any reason, the event
$C\Rightarrow c$ doesn't have the given probability density, it
does not means that the theory is wrong but it suggests that,
perhaps, $\varrho(a)$ and $\pi(b)$ are not the probability
densities of the random variables $A$ and $B$.
\section{QUANTUM RANDOMNESS}
We will prove the following theorem. Assume two observable of a
quantum system represented by the operators $A$ and $B$ in a
Hilbert space. Let $\varrho(a)$ and $\pi(b)$ be the distributions
of the eigenvalues of the operators as predicted by quantum
mechanics for the system in a state $\Psi$, that is,
$\varrho(a)=\langle\Psi,P_{a}\Psi\rangle$ where $P_{a}$ is the
projector in the invariant subspace of the eigenvalue $a$ (in the
case of nondegeneracy of the eigenvalue $a$ with eigenvector
$\varphi_{a}$, we have
$P_{a}=\varphi_{a}\langle\varphi_{a},\bullet\rangle$ and we get
the more familiar expression
$\varrho(a)=|\langle\varphi_{a},\Psi\rangle|^{2}$) and similarly,
$\pi(b)=\langle\Psi,P_{b}\Psi\rangle$
($\pi(b)=|\langle\phi_{b},\Psi\rangle|^{2}$ if $b$ is
nondegenerate with eigenvector $\phi_{b}$). Let $C=F(A,B)$ be an
observable built as a function of the observables $A$ and $B$
whose operator has  the eigenvalues $c$ and the corresponding
projector $P_{c}$ (eigenvector $\xi_{c}$ if nondegenerate). Then,
considering the quantum mechanical distribution $\sigma_{QM}(c)$
and the probability density provided by the theory of random
variables $\sigma_{RV}(c)$ given by
\begin{eqnarray}
  \sigma_{QM}(c) &=& \langle\Psi,P_{c}\Psi\rangle \\
  \sigma_{RV}(c) &=& \int\!\!\!da\!\!\!\int\!\!\!db\ \varrho(a)\ \pi(b)
    \ \delta\left(c-F(a,b)\right)\ ,
\end{eqnarray}
we can prove the following:
\begin{enumerate}
    \item if $[A,B]\neq 0$ then in general it is
    $\sigma_{QM}(c)\neq\sigma_{RV}(c)$;
    \item if $[A,B]=0$ then we have two cases:\begin{enumerate}
        \item if they commute because they are related by a
        function like $B=G(A)$, then
        $\sigma_{QM}(c)=\sigma_{RV}(c)$;
        \item if they commute because they are independent
        observables related to different degrees of freedom, then
        we have again two cases:\begin{enumerate}
            \item if the state is factorizable in the product
            Hilbert space corresponding to the different degrees
            of freedom, then $\sigma_{QM}(c)=\sigma_{RV}(c)$;
            \item if the state is not factorizable (entangled) in the product
            Hilbert space corresponding to the different degrees
            of freedom, then $\sigma_{QM}(c)\neq\sigma_{RV}(c)$.
        \end{enumerate}
    \end{enumerate}
\end{enumerate}
Summarizing we have
\begin{equation}
\left\{%
\begin{array}{ll}
  \ [A,B] \neq 0 , & \ \ \ \ \ \sigma_{QM}(c)\neq \sigma_{RV}(c); \\
  \ [A,B]=0 , & \left \{
\begin{array}{ll}
    B=G(A), & \ \ \ \ \sigma_{QM}(c)=\sigma_{RV}(c); \\
    A \otimes \textbf{1},\textbf{1}\otimes B, & \left\{
\begin{array}{ll}
    \Psi=\chi\otimes \psi, & \sigma_{QM}(c)=\sigma_{RV}(c); \\
    \Psi \neq \chi \otimes \psi, & \sigma_{QM}(c)\neq \sigma_{RV}(c). \\
\end{array}
\right.     \\
\end{array}
\right.     \\
\end{array}
\right.
\end{equation}
That is, in general the quantum mechanical distribution and the
random variables densities are different and only in some cases
(commuting operators and factorizable state) they can be equal.

In order to compare them, we can write Eqs.(3) and (4) in a way
to emphasize their difference (assuming nondegenerate eigenvalues
$a,b$ and $c$) as
\begin{eqnarray}
  \sigma_{QM}(c) &=& \int\!\!\!da\!\!\!\int\!\!\!db\ \langle\varphi_{a},\Psi\rangle
  \langle \Psi, \phi_{b}\rangle \left\langle\phi_{b},
  \ \ \ \ \ \ \xi_{c}\langle\xi_{c},\bullet\rangle\ \ \ \ \ \ \varphi_{a}\right\rangle
    \ \\
  \sigma_{RV}(c) &=& \int\!\!\!da\!\!\!\int\!\!\!db\ \langle\varphi_{a},\Psi\rangle
  \langle \Psi, \phi_{b}\rangle \left\langle\phi_{b},
  \ \ \ \delta\left(c-F(a,b)\right)\Psi\langle\Psi,\bullet\rangle\ \ \ \varphi_{a}\right\rangle
    \ \ ,
\end{eqnarray}
The QM case involves the projector in the state $\xi_{c}$ whereas
 in the RV case it involves the projector in the state $\Psi$ and
a constraint $\delta\left(c-F(a,b)\right)$. Of course this formal
difference only suggests that $\sigma_{QM}\neq \sigma_{RV}$ but
it does not prove it because the different operators are inside
an integral. We will later prove anyway that they are indeed
different. A nice comparisons of the two distributions (3) and
(4) is obtained when we calculate the moments of the
distributions. For the QM case we have
\begin{equation}\label{EnQM}
 E^{(n)}_{QM}= \int\!\!\!dc\ c^{n}\sigma_{QM}(c) =
 \left\langle\Psi, \int\!\!\!dc\ c^{n}P_{c}\ \Psi \right\rangle
 =\left\langle\Psi, C^{n}\Psi \right\rangle\ =
\left\langle\Psi, F^{n}(A,B)\Psi \right\rangle\\ ,
\end{equation}
and in the RV case we have
\begin{equation}\label{EnRV}
 E^{(n)}_{RV}= \int\!\!\!dc\ c^{n}\sigma_{RV}(c) =
 \int\!\!\!dc\ c^{n} \int\!\!\!da\!\!\!\int\!\!\!db\ \varrho(a)\ \pi(b)
    \ \delta\left(c-F(a,b)\right) =
    \int\!\!\!da\!\!\!\int\!\!\!db\  F^{n}(a,b)\varrho(a)\ \pi(b)
 \ .
\end{equation}
Inserting the completeness relation in Eq.(\ref{EnQM}) and
replacing the quantum mechanic expression for the densities
$\varrho$ and $\pi$ we get
\begin{eqnarray}
  E^{(n)}_{QM} &=& \int\!\!\!da\!\!\!\int\!\!\!db\ \left\langle\Psi
  ,P_{a}F^{n}(A,B)P_{b}\ \Psi\right\rangle \ ,\\
  E^{(n)}_{RV} &=& \int\!\!\!da\!\!\!\int\!\!\!db\ \left\langle\Psi
  ,P_{a}F^{n}(a,b)P_{\Psi}P_{b}\ \Psi\right\rangle \ .
\end{eqnarray}
Notice that the QM case involves the \emph{operator} $F^{n}(A,B)$
and the RV case the \emph{number} $F^{n}(a,b)$ multiplied by the
projector in the state $\Psi$. A nicer comparison is obtained in
the particular case where $[A,B]=0$. In this case we can write
$F^{n}(A,B)$ as a power series with the operator $B$ always to
the right and $A$ to the left and using the relations
$BP_{b}=bP_{b}$ and $P_{a}A=P_{a}a$ we have
\begin{equation}\label{EnQMRV}
   [A,B]=0\rightarrow
   \left\{%
\begin{array}{ll}
    E^{(n)}_{QM} &=  \int\!\!da\int\!\!db\ F^{n}(a,b) \left\langle\Psi
  ,\ \ P_{a}P_{b}\ \ \Psi\right\rangle
  = \int\!\!da\int\!\!db\ F^{n}(a,b)\ \ \ \texttt{Tr}\left\{
   P_{a}P_{b}P_{\Psi}\right\}\ , \\
    E^{(n)}_{RV} &=  \int\!\!da\int\!\!db\ F^{n}(a,b)\left\langle\Psi
  ,P_{a}P_{\Psi}P_{b}\ \Psi\right\rangle
  =\int\!\!da\int\!\!db\ F^{n}(a,b)\ \texttt{Tr}\left\{
   P_{a}P_{\Psi}P_{b}P_{\Psi}\right\}\ ,
\end{array}%
\right.
\end{equation}
where we have used the trace expression for the expectation
values. Clearly the moments, and therefore the distributions, are
different in general although for some states they may be equal
(for instance when $P_{b}$ and $P_{\Psi}$ commute).

 Considering some examples it is clear that for
noncommuting observables, like position  and momentum, it must be
$\sigma_{QM}\neq \sigma_{RV}$. Take for instance $A=X^{2},
B=P^{2}$ and $C=X^{2}+P^{2}$ where $\varrho(a)$ and $\pi(b)$ are
continuous, and therefore $\sigma_{RV}(c)$ is also continuous,
whereas $\sigma_{QM}(c)$ is not continuous because it corresponds
to the energy distribution of an harmonic oscillator.

Another case  of physical interest with noncommuting observables
is when $A=X$, the position of a free particle at $t=0$,
$B=\frac{t}{m}P$, its velocity multiplied by the time $t$, and
$C=A+B=X+P\frac{t}{m}$ corresponds to the position of the
particle at time $t$. The quantum mechanical prediction for the
distribution of position at time $t$ clearly differs from the
random variable distribution. To see this we can, for instance,
calculate the width of both distributions
$\Delta^{2}=E^{(2)}-(E^{(1)})^{2}$ using Eqs.(8,9), or directly
from Eqs. (3,4), and we obtain
\begin{eqnarray}
  \Delta^{2}_{QM} &=& \Delta^{2}_{x} +\frac{t^{2}}{m^{2}}\Delta^{2}_{p}+
   \frac{t}{m}\left( \langle \Psi,(XP+PX)\Psi\rangle-2\langle \Psi, X\Psi\rangle
   \langle\Psi, P\Psi\rangle\right)\ ,\\
  \Delta^{2}_{RV} &=& \Delta^{2}_{x} +\frac{t^{2}}{m^{2}}\Delta^{2}_{p}\
  ,
\end{eqnarray}
where  $\Delta_{x}$ and $\Delta_{p}$ are the width of the
probability densities of position and momentum $\varrho(x)$ and
$\pi(p)$. The conceptual relevance of this difference will be
discussed later. Notice that even if the observables $A$ and $B$
would commute, we would still get different results in some
states where the corresponding correlation term in Eq.(13) does
not vanish.

Let us consider now the case  of $[A,B]=0$ because $B=G(A)$. In
this case, if we know the density $\varrho(a)$ for the random
variable $A$ then we calculate the density of $B$ using
Eq.(\ref{functrv}) as
\begin{equation}\label{BGA}
   \pi(b) = \int\!\!\!da\ \varrho(a)\ \delta\left(b-G(a)\right)\
   .
\end{equation}
This result of the theory of random variables is also obtained in
quantum mechanics where $\pi(b) = \langle\Psi,P_{b}\Psi\rangle $.
If $[A,B]=0$, the operators A and $B$ share eigenvectors, that
is, they have the same projectors. However the function $G(A)$
can introduce more degeneracy because there may be different
values $a \neq a'$ with $G(a)=G(a')$ corresponding to the same
eigenvalue $b$ (think for instance that for $X^{2}$ we have
$P_{x^{2}}=P_{x}+P_{-x}$). Therefore the projector $P_{b}$ is
obtained adding all projectors $P_{a}$ where $a$ and $b$ satisfy
the condition $b=G(a)$. That is,
\begin{equation}\label{BGA1}
   P_{b} = \int\!\!\!da\ P_{a}\ \delta\left(b-G(a)\right)\
   .
\end{equation}
Now, taking the expectation value $\langle\Psi,P_{b}\Psi\rangle$
we immediately obtain the relation in Eq.(\ref{BGA}). In this
case, the observable $C=F(A,B)$ does not really depends on $A$
and $B$ but is a function of just one observable $C=F(A,G(A))$
and repeating the same argument it is clear that the theory of
random variables and quantum mechanics predict the same density:
$\langle\Psi,P_{c}\Psi\rangle = \int\!\!da\ \varrho(a)\
\delta\left(c-F(a,G(a))\right)$.

We can now analyse the case where the observables $A$ and $B$
commute because they correspond to different degrees of freedom.
These operators act in different Hilbert spaces $\mathcal{H}_{1}$
and $\mathcal{H}_{2}$ and the system is described by a state in
the space $\mathcal{H}=\mathcal{H}_{1}\otimes\mathcal{H}_{2}$.
Given a state $\Psi$, there is a basis $\{\chi_{k}\}$ in
$\mathcal{H}_{1}$ and a basis $\{\zeta_{k}\}$ in
$\mathcal{H}_{2}$ that allow the \emph{bi-orthogonal}
decomposition of the state
\begin{equation}\label{biort}
   \Psi = \sum_{k=1}^{N} \alpha_{k}\ \chi_{k}\otimes\zeta_{k}\ .
\end{equation}
As is well known, if $N=1$ the state is factorizable and if
$N\geq 2$ the state is entangled or nonfactorizable. The
operators and their corresponding projectors are extended in the
product space as $A\otimes\textbf{1},
P_{a}\otimes\textbf{1},\textbf{1} \otimes B, \textbf{1} \otimes
P_{b}$, and the densities associated with the observables are
given by quantum mechanics as
$\varrho(a)=\langle\Psi,P_{a}\otimes\textbf{1}\Psi\rangle$ and
$\pi(b)=\langle\Psi,\textbf{1}\otimes P_{b}\Psi\rangle$. The
operator $C=F(A,B)$ will have projectors $P_{c}$ associated with
the eigenvalues $c$ that project in a subspace of
$\mathcal{H}_{1}\otimes\mathcal{H}_{2}$. We will prove that
\begin{equation}\label{projc}
    P_{c}= \int\!\!\!da\!\!\!\int\!\!\!db\
    \delta\left(c-F(a,b)\right)\ P_{a}\otimes P_{b}\ .
\end{equation}
In order to prove it, let us consider the function $F(A,B)$
expanded as a formal power series
\begin{equation}\label{projc1}
   C=F(A,B)=\sum_{n,m} c_{n,m} A^{n}\otimes B^{m}\ .
\end{equation}
Now, from the spectral decomposition of $A=\int\!da\ a\ P_{a}$
and using the orthogonality and idempotent property of the
projectors $P_{a}P_{a'}=\delta(a-a')P_{a}$ we easily find that
$A^{n}=\int\!da\ a^{n}\ P_{a}$ and similarly $B^{m}=\int\!db\
b^{m}\ P_{b}$. Replacing above and reconstructing the function we
get
\begin{equation}\label{projc2}
   C=F(A,B)=\int\!\!\!da\!\!\!\int\!\!\!db\
    F(a,b)\ P_{a}\otimes P_{b}\ .
\end{equation}
Now we write the function $F(a,b)$ as an integral over $c$ with a
Dirac distribution
\begin{equation}\label{projc2}
  C= F(A,B)=\int\!\!\!da\!\!\!\int\!\!\!db\
    \int\!\!\!dc\ c\ \delta\left(c-F(a,b)
   \right)\ P_{a}\otimes P_{b} =
  \int\!\!\!dc\ c\ \int\!\!\!da\!\!\!\int\!\!\!db\
    \delta\left(c-F(a,b)
   \right)\ P_{a}\otimes P_{b} \ .
\end{equation}
But this is precisely the spectral decomposition of the operator
$C$ and therefore the double integral over $a$ and $b$ is the
projector $P_{c}$ as given in Eq.(\ref{projc}). We obtain the
quantum mechanical distribution of the eigenvalues of the
operator $C$ by taking the expectation value in Eq.(\ref{projc})
\begin{equation}\label{sigmqc}
    \sigma_{QM}(c)=\langle\Psi, P_{c}\Psi\rangle= \int\!\!\!da\!\!\!\int\!\!\!db\
    \delta\left(c-F(a,b)\right)\ \langle\Psi, P_{a}\otimes P_{b}\Psi\rangle\ .
\end{equation}
In this expression we consider now the two cases for the state,
factorizable or entangled. In the first case the state is given
by $\Psi=\chi\otimes\zeta$ and inserting above we have
\begin{equation}\label{sigmqc1}
    \sigma_{QM}(c)= \int\!\!\!da\!\!\!\int\!\!\!db\
    \delta\left(c-F(a,b)\right)\ \langle \chi, P_{a}\chi\rangle
    \langle\zeta, P_{b}\zeta\rangle\ ,
\end{equation}
but $\langle \chi, P_{a}\chi\rangle=\langle \Psi,
P_{a}\otimes\textbf{1}\Psi\rangle=\varrho(a)$ and $\langle\zeta,
P_{b}\zeta\rangle=\pi(b)$ and therefore
\begin{equation}\label{sigmqc2}
    \sigma_{QM}(c)= \int\!\!\!da\!\!\!\int\!\!\!db\
    \delta\left(c-F(a,b)\right)\ \varrho(a)\ \pi(b)\ ,
\end{equation}
which is the density predicted by the theory of random variables;
that is, $\sigma_{QM}(c)=\sigma_{RV}(c)$. In the non-factorizable
case we have
\begin{equation}\label{sigmqc3}
    \sigma_{QM}(c)= \int\!\!\!da\!\!\!\int\!\!\!db\
    \delta\left(c-F(a,b)\right)\ \sum_{k,r}\alpha^{\ast}_{k}\alpha_{r}
    \langle \chi_{k}, P_{a}\chi_{r}\rangle
    \langle\zeta_{k}, P_{b}\zeta_{r}\rangle\ ,
\end{equation}
which contains all non-diagonal terms. Notice that here
$\varrho(a)=\sum_{k}|\alpha_{k}|^{2}\langle \chi_{k},
P_{a}\chi_{k}\rangle$ and $\pi(b)=\sum_{k}|\alpha_{k}|^{2}\langle
\zeta_{k}, P_{b}\zeta_{k}\rangle$ and then
\begin{equation}\label{sigmqc4}
    \sigma_{RV}(c)= \int\!\!\!da\!\!\!\int\!\!\!db\
    \delta\left(c-F(a,b)\right)\
    \sum_{k,r}|\alpha_{k}|^{2}|\alpha_{r}|^{2}
    \langle \chi_{k}, P_{a}\chi_{k}\rangle
    \langle\zeta_{r}, P_{b}\zeta_{r}\rangle\ ,
\end{equation}
therefore in general we have $\sigma_{QM}(c)\neq\sigma_{RV}(c)$.
\section{CONCLUSION}
The general conclusion that we can draw from this study is that
quantum mechanics is not a random process in space. If it where,
then the randomness in quantum mechanics should be described by
the theory of random variables and we have seen that this is not
the case.

In particular, we saw in the last section that if we know the
probability density for the position of a particle $\varrho(x)$
at one time, and we know the probability density of its velocity
(or momentum) $\pi(p)$, then, according to quantum mechanics, the
probability density for the position at a later time \emph{is
not} calculated by sampling one position and adding a
displacement sampled from the velocity distribution, as would be
calculated with the theory of random variables. Notice that this
result contradicts the gnoseological interpretation of the
probability for position and velocity of the particle. Indeed, if
the particle \emph{has a definite position}, unknown to us, with
its value distributed according to $\varrho(x)$, and the particle
\emph{has a definite velocity}, unknown to us and distritbuted
according to  $\pi(p)$, then the position at a later time
\emph{must} given by the rule ``$x_{0}+vt$'' where $x_{0}$ and
$v$ are taken from their distributions. In other words, the
location at time $t$ \emph{must} be given according to the theory
of random variables; but this is contrary to the quantum
mechanical result! This can only mean that $X$ and $P$ are not
random variables and $\varrho(x)$ and $\pi(p)$ are not densities
of probabilities although they are, in principle, measured as if
they were, that is, by counting the frequency of realization of
the events $X\Rightarrow x$ and $P\Rightarrow p$. To call
``probability'' to something that is not a probability, that is,
something that does not complies with the rules  of
probabilities, is perhaps a misuse of language and it could be
convenient to denote the quantum mechanical distributions with
another name. It is unfortunate that the term ``probability'' is
irreversibly installed in quantum mechanics because, strictly
speaking, they are not probabilities and the misnomer introduces
confusion, not only in the teaching, but also in search of an
interpretation of quantum mechanics. A more appropriated name for
$\varrho(x)$ and $\pi(p)$ could be for instance the
\emph{existential weight} of the events $X\Rightarrow x$ and
$P\Rightarrow p$ in the system in a particular state. Although
these existential weights are measured in an experiment in the
same way that one measures a probability, these quantities are
not probabilities and do not necessarily obey all the rules of
probabilities dictated by the theory of random variables.
Probabilities propagate according to the theory of random
variables but the existential weights are calculated by the rules
of quantum mechanics. Accordingly we can think about the position
of a particle as a diffuse observable without a putative value.
Every possible value, every event $X\Rightarrow x$, has a
propensity to appear in an observation given by the existential
weight, but this does not means that the event has that value
with some probability and that the measurement of the observable
exhibits a pre-existent value. This idea has been synthetically
expressed by A. Peres saying that ``unperformed experiments have
no result''\cite{per}.

The results presented in this work suggest that we can not
consider $\varrho(x)$ and $\pi(p)$ to be the probability
densities of two independent random variables associated with the
position and momentum of a particle. We should consider now the
possibility that position and momentum \emph{are not independent}
and are related by some unknown connection that is responsible
for the quantum mechanical correlations. We could then recover
the theory of random variables for quantum mechanics if we could
show that $\varrho(x)$ and $\pi(p)$ are the marginal
distributions of a \emph{joint probability density} $W(x,p)$.
There are many possibilities to achieve this, the most famous is
the Wigner phase space distribution, but it was shown\cite{cohen}
that the most general function producing the same expectation
values as quantum mechanics is not nonnegative everywhere and
therefore it can not be considered as a genuine joint probability
density.
\section{Acknowledgements}
This work received partial support from ``Consejo Nacional de
Investigaciones Cient{\'\i}ficas y T{\'e}cnicas'' (CONICET), Argentina.

\end{document}